\documentclass[preprint,proceedings]{rmaa}
\usepackage{paralist}

\usepackage{psfrag,color}

\def\mabs{$M_{\rm B}$}

\def\hii{H{\sc ii}}

\def\doh{$12 + \log(\rm O/H)$}
\def\lno{$\log(\rm N/O)$}
\def\kms{km s$^{-1}$}

\def\msun{M$_{\odot}$}

\def\zsun{Z$_{\odot}$}
\def\halpha{\ifmmode {\rm H{\alpha}} \else $\rm H{\alpha}$\fi}
\def\hbeta{\ifmmode {\rm H{\beta}} \else $\rm H{\beta}$\fi}
\def\oii{[O\,{\sc ii}] $\lambda\lambda$3726,3728}

\def\oiii{[O\,{\sc iii}] $\lambda\lambda$4959,5007}

\def\nii{[N\,{\sc ii}] $\lambda$6584}

\def\civ{C\,{\sc iv} $\lambda$1550}
\def\siii{Si\,{\sc iv} $\lambda\lambda$1527,1533}
\def\rr23{$R_{\rm 23}$}
\def\oo32{$O_{\rm 32}$}
\def\xx3{$X^{*}_{\rm 3}$}
\newcommand{\ltapprox}{\raisebox{-0.5ex}{$\,\stackrel{<}{\scriptstyle\sim}\,$}}
\newcommand{\gtapprox}{\raisebox{-0.5ex}{$\,\stackrel{>}{\scriptstyle\sim}\,$}}

\SetYear{2004}
\SetConfTitle{Second International Workshop on Science with the GTC}

\title{Star-Forming Galaxies at $z \sim 2$: a major science case for the EMIR/GOYA survey on GTC} 

\author{
  T. Contini,\altaffilmark{1} 
  M. Lemoine-Busserolle,\altaffilmark{2}
  R. Pell\' o,\altaffilmark{1}
  J.-F. Le Borgne,\altaffilmark{1}
  and J.-P. Kneib\altaffilmark{3,1}}

\altaffiltext{1}{Laboratoire d'Astrophysique de Toulouse et Tarbes, Toulouse, France}
\altaffiltext{2}{Institute of Astronomy, Cambridge, UK}
\altaffiltext{3}{California Institute of Technology, Pasadena, USA}
\shortauthor{T. Contini et al.}
\shorttitle{Star-Forming Galaxies at $z \sim 2$}

\fulladdresses{
\item T. Contini, R. Pell\' o, and J.-F. Le Borgne: Laboratoire d'Astrophysique de Toulouse et Tarbes - UMR 5572, Observatoire Midi-Pyr\'en\'ees, 14 avenue E. Belin, F-31400 Toulouse, France (\email{contini, roser, leborgne@ast.obs-mip.fr}).
\item J.-P. Kneib: California Institute of Technology, Pasadena, CA 91125, USA (\email{kneib@caltech.edu}).
\item M. Lemoine-Busserolle: Institute of Astronomy, University of Cambridge, Madingley Road, Cambridge, CB4 OH4, UK (\email{lemoine@ast.cam.ac.uk}).
}

\listofauthors{T. Contini, M. Lemoine-Busserolle, R. Pell\' o, J.-F. Le Borgne, \& J.-P. Kneib}
\indexauthor{Contini, T.}
\indexauthor{Lemoine-Busserolle, M.}
\indexauthor{Pell\' o, R.}
\indexauthor{Le Borgne, J.-F.}
\indexauthor{Kneib, J.-P.}

\abstract{We present the first results of a project aiming to derive the physical 
properties of high-redshift lensed galaxies, intrinsically fainter than the 
Lyman break galaxies currently observed in the field. From FORS and ISAAC 
spectroscopy on the VLT, we use the full rest-frame UV-to-optical range to 
derive the physical properties (SFR, extinction, chemical abundances, dynamics, 
mass, etc) of low-luminosity $z \sim 2$ star-forming galaxies. Although the sample 
is still too small for statistical studies, these results give an insight into 
the nature and evolutionary status of distant star-forming objects and their 
link with present-day galaxies. Such a project will serve as a basis for the 
scientific analysis of the EMIR/GOYA survey on the GTC.}

\addkeyword{galaxies: evolution}
\addkeyword{galaxies: starburst}
\addkeyword{galaxies: abundances}
\addkeyword{galaxies: infrared}
\addkeyword{galaxies: kinematics and dynamics}

\begin{document}
% Typeset article header
\maketitle

\section{Motivations}
\label{sec:intro}

Obtaining statistically significant samples of galaxies, from the local 
Universe to the highest redshifts, is mandatory to constrain the models 
of galaxy formation and evolution. Large spectroscopic samples of galaxies 
at all redshifts have become available during the last ten years, thanks to 
massive surveys in the different rest-frame wavelength domains 
(e.g. Lilly et al. 1995; Steidel et al. 1996, 1999, 2004; 
Colless et al. 2001; Pettini et al. 2001; Schneider et al. 2003). 
The evolution of the physical properties of 
galaxies as a function of redshift is particularly important for 
galaxies at z \gtapprox 1, a redshift domain where galaxies are expected 
to be strongly affected by merging or assembly processes. 

In this context, massive clusters acting as Gravitational Telescopes (GTs) 
constitute a powerful tool to study the high-redshift Universe. They have been
successfully used over a wide range of wavelengths, ranging from the UV 
to the sub-mm (e.g. Pell\'o et al.\ 2003 and references therein) 
and allowed recently the detection of the most distant galaxies 
known so far ($z \ga 7$; Pello et al.\ 2004; Kneib et al.\ 2004; see also the 
contribution by R. Pello in these proceedings). The large 
magnification factors of galaxies that are close to the critical lines 
(typically 1 to 3 magnitudes) can be used to probe the physical properties 
of {\em intrinsically faint} high-redshift galaxies, which would otherwise be 
beyond the limits of conventional spectroscopy (e.g. Ebbels et al. 1996; 
Pell\'o et al. 1999, Pettini et al.\ 2000, Mehlert et al. 2001). 

The recent development of near-IR (NIR) spectrographs on 10m-class
telescopes has allowed the study of the rest-frame optical properties
of high-redshift galaxies. Indeed, NIR spectroscopy allows to
access the most relevant optical emission lines (\halpha, \hbeta, \oii, 
\oiii, etc), in order to probe the physical properties of galaxies (SFR,
reddening, chemical abundances, kinematics, virial mass, etc), all the way
from the local universe to $z \sim 4$, using the same parameter space
and indicators. The pioneering work by Pettini and collaborators 
(1998, 2001) has shown that the rest-frame optical properties of the 
{\em brightest} Lyman break galaxies (LBGs) at $z \sim 3$ are relatively 
uniform. A new sample of 16 LBGs at $2 \le z \le 2.6$ has been recently 
published by Erb et al. (2003) and, in this case, significant differences 
are found in the kinematics of galaxies at $z \sim 2$ compared to $z \sim 3$. 

However, the current studies of the high-redshift Universe are limited to 
the most luminous/massive galaxies (\mabs $< -22$). Moreover, the 
$1.5 \ltapprox z \ltapprox 2.5$ redshift interval (the so-called 
``redshift desert'') remains relatively poorly known because of the lack 
of strong spectral features that can be used to identify sources at such 
redshift. We thus started, about two years ago, an observational program 
to perform a detailed study of the physical properties for a significant 
sample of amplified high-redshift galaxies by using the full rest-frame UV/optical 
spectral features (observed in the optical/NIR domains respectively) in order 
to build a comprehensive picture of the nature of the low-luminosity galaxy 
population at $z \geq 1.5$.

In this paper, we present the first results obtained on three galaxies at $z \sim 2$ 
from our ongoing spectroscopic program with FORS and ISAAC on the VLT 
(Lemoine-Busserolle et al. 2003, 2004; Le Borgne et al. 2004). Our ultimate 
objective is to obtain rest-frame UV and optical spectroscopy for a sample of 
$\sim$ 15 gravitationally amplified high-redshift galaxies to study their 
physical properties. This project can be considered as 
a pilot program for the future NIR multi-object spectrograph EMIR on GTC.

\section{Physical properties from the rest-frame UV-to-optical spectra}
\label{sec:data}

\begin{figure*}[t]
  \includegraphics[width=\columnwidth,height=8cm]{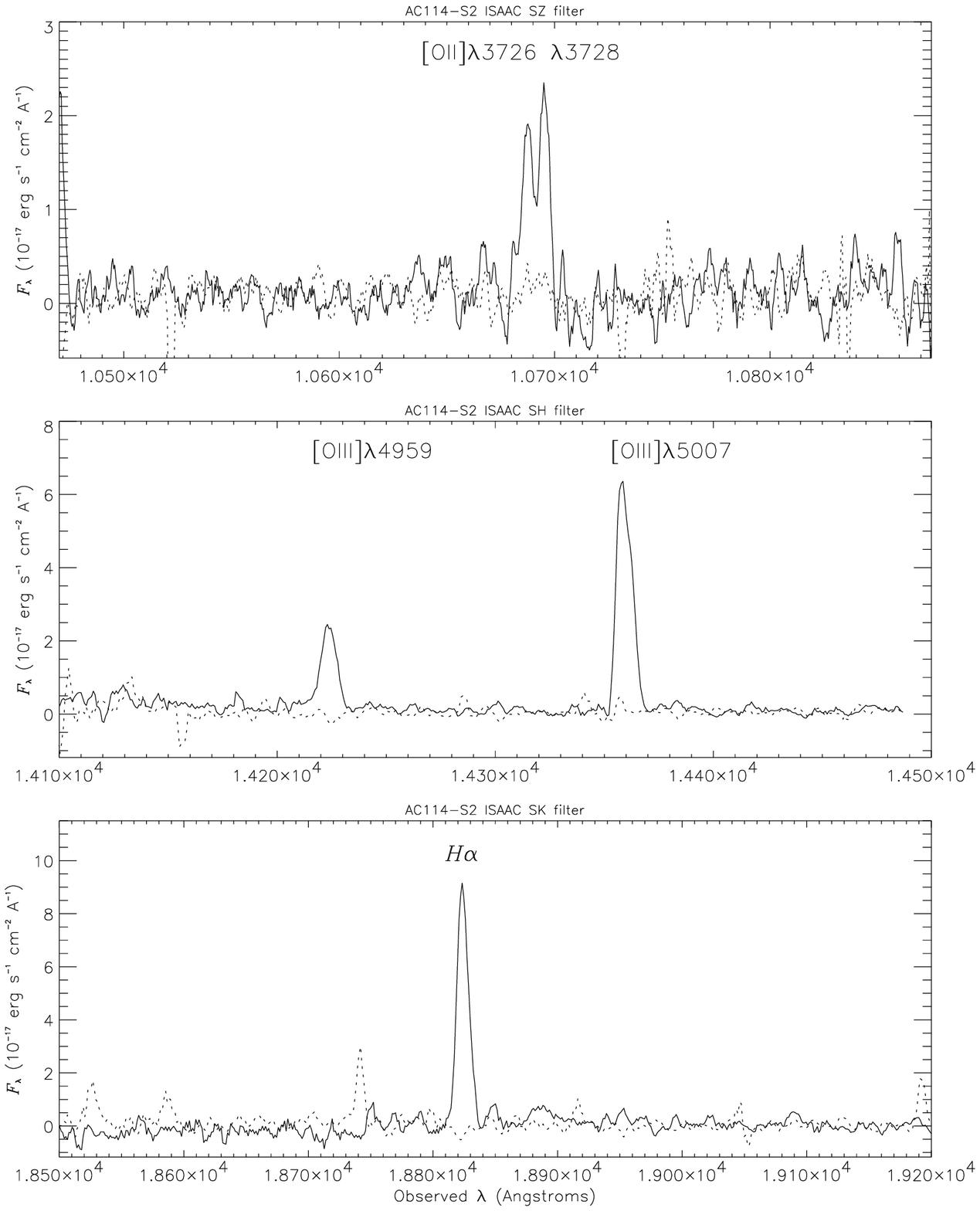}%
  \hspace*{\columnsep}%
  \includegraphics[width=\columnwidth,height=8cm]{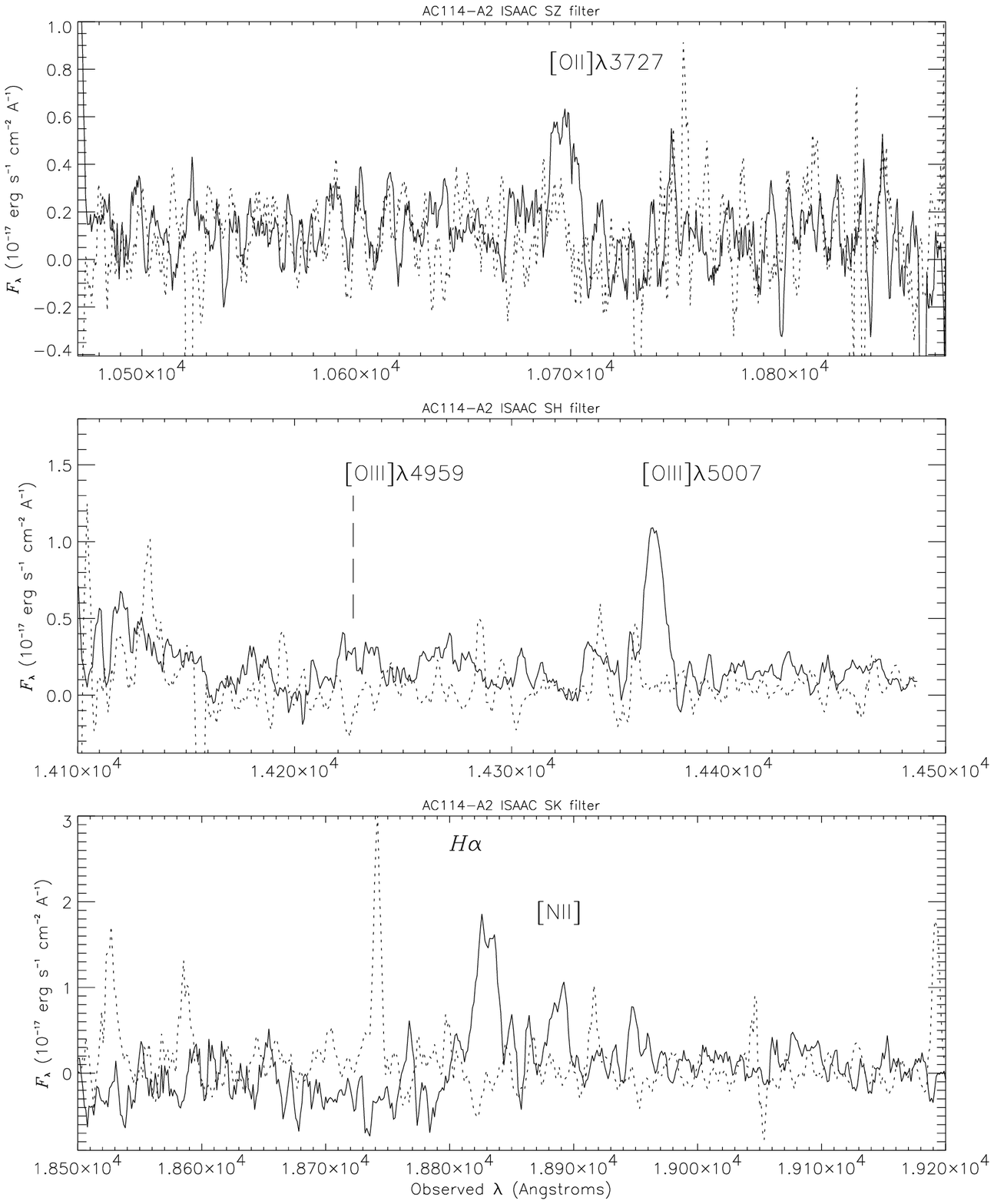}
  \caption{The extracted one-dimensional $Y$, $H$, and $K$ band 
spectra of AC114$-$S2 (left) and AC114$-$A2 (right) taken with the ISAAC 
spectrograph on the VLT. The dashed line shows the 1 $\sigma$ error spectrum. 
Spectra have been smoothed to the instrumental resolution.
}
  \label{fig:spectra}
\end{figure*}

Both rest-frame UV (Leborgne et al. 2004; Lemoine-Busserolle et al. 2004) 
and rest-frame optical (Lemoine-Busserolle et al. 2003, 2004) spectra 
have been obtained on three lensed galaxies at $z \sim 2$ using FORS and 
ISAAC respectively on the VLT. 

Examples of $YHK$-bands ISAAC spectra are shown 
in Fig.~\ref{fig:spectra} for the $z=1.87$ lensed galaxies S2 and A2 
located behind the massive cluster AC 114 ($z=0.31$). 
Relatively long integration times (typically 
3-5 hours per band) are mandatory to get high enough S/N ratios. The brightest 
rest-frame optical emission lines (\oii, \oiii, and \halpha) are clearly 
detected and measured in these spectra, as well as in the spectrum of 
the giant blue arc ($z=2.24$, hereafter Cl2244-gA) located behind the 
lensing cluster Cl 2244-02 ($z=0.33$). 
\hbeta\ emission-line is not seen in these spectra as it 
falls at the blue edge of the $H$ band. The \nii\ is detected only in the 
spectrum of A2; an upper limit for the flux of this line has 
been derived for A2 and Cl2244-gA. Measuring this nitrogen line is very 
important to break the degeneracy between the oxygen abundance O/H and 
the strong emission-line intensity in order to get an accurate 
value of the metallicity. The analysis of rest-frame optical 
emission lines is fully described in Lemoine-Busserolle et al. (2003). 
We have been able to derive accurate measurements of the oxygen 
abundance O/H (as indicator of metallicity), the nitrogen-to-oxygen 
abundance ratio N/O, the SFR from \halpha\ luminosities, and the 
virial masses from the FWHM of the brightest emission lines (\oiii, and 
\halpha). Moreover, thanks to the gravitational amplification, the line 
profiles of S2 are spatially resolved, leading to a velocity gradient of 
$\pm 240$ \kms, which yields a dynamical mass for this galaxy of $\sim 
1.3 \times 10^{10}$ \msun\ within the inner 1 kpc radius 
(Lemoine-Busserolle et al. 2003).

The rest-frame UV spectra of two lensed galaxies are shown in 
Leborgne et al. (2004) for S2 and Lemoine-Busserolle et al. (2004) 
for Cl2244-gA. These spectra have been mainly used to constrain 
i) the age od the star-forming events by a detailed model fit of 
the \civ\ and \siii\ lines, ii) the overall dust extinction using 
the observed SED (rest-frame UV-to-optical range) and the latest 
version of the Bruzual \& Charlot (2003) population synthesis models, 
and iii) an independant estimate of the (stellar) metallicity using 
both the UV continuum slope and the strength of the interstellar 
absorption lines. 

All the relevant physical parameters of the three lensed 
galaxies, derived from both the rest-frame UV and optical spectra, 
are summarized in Table~\ref{tab:param}.

\begin{table*}[!t]\centering
  \tablecols{9}
  % Stretch the space between table columns 
%  \setlength{\tabcolsep}{2.8\tabcolsep}
  \caption{Physical parameters of $z \sim 2$ lensed galaxies} \label{tab:param}
\begin{tabular}{lcccccrrr}
    \toprule
Galaxy & $z$ & \mabs & \doh & \lno & E(B-V) & Age  & $\sigma$  & $M_{\rm vir}$ \\
 &  &  &  &  & [mag] & [Myr] & [\kms] & [$10^{10}$ \msun]\\
    \midrule
AC114-S2 & 1.87 & $-20.60$& $7.25\pm 0.2$ & $< -1.32$ & 0.2 & $7\pm 2$ & $55\pm 6$ & $0.53\pm 0.12$ \\
AC114-A2 & 1.87 & $-19.80$& $8.94\pm 0.2$ & $-0.42\pm 0.2$ & 0.4 & {\nodata} & $105\pm 15$ & $2.36 \pm 0.67$ \\
Cl2244-gA & 2.24 & $-21.80$& $8.27\pm 0.4$ & $< -0.68$ & 0.0 & $10 - 15$ & $32\pm 8$ & ($0.32 \pm 0.16$) \\
    \bottomrule
\end{tabular}
\end{table*}

\section{The nature of high-z galaxies}
\label{sec:anal}

\begin{figure}[!t]
  \includegraphics[width=\columnwidth]{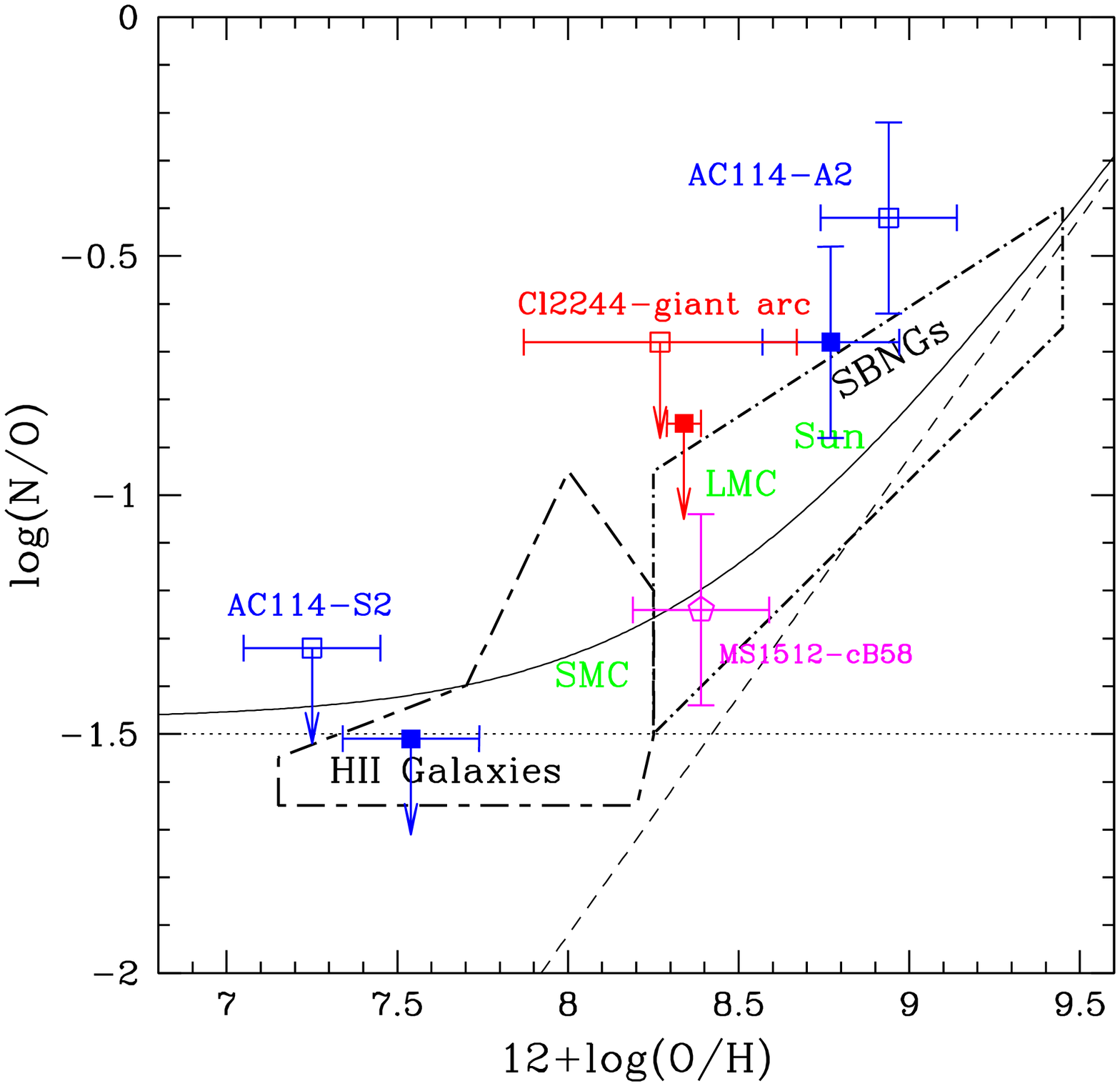}
  \caption{Nitrogen-to-oxygen abundance ratio as a function
of oxygen abundance. The location of the three $z \sim 2$ 
galaxies: S2, A2 in AC 114 and the giant blue arc in Cl 2244, is 
shown with (filled squares) and without extinction correction 
(empty squares). 
Two comparison samples of nearby star-forming 
galaxies are shown (see Contini et al. 2002 for references): 
starburst nucleus galaxies (dot -- short dash line) and \hii\ galaxies 
(short dash -- long dash line). The position of the Sun, LMC, SMC, and 
the bright lensed galaxy MS1512-cB58 is also indicated. 
%We show, for comparison, samples of 
%intermediate-redshift galaxies: UV-selected galaxies at 
%$z \sim 0.1-0.4$ (circles; Contini et al. 2002), and emission-line 
%galaxies at $z \sim 0.1-0.5$ (stars; Kobulnicky \& Zaritsky 1999). 
%The location of Damped Ly$\alpha$ systems (triangles, filled symbols 
%are upper limits; Pettini et al. 2002b) is shown for comparison. 
%Theoretical curves for a {\em primary} (dotted line), a 
%{\em secondary} (dashed line), and a {\em primary + secondary} 
%(solid line) production of nitrogen (Vila-Costas \& Edmunds 
%1993) are shown.
}
  \label{fig:novsoh}
\end{figure}

In Figure~\ref{fig:novsoh}, we examine the location of high-redshift 
galaxies in the N/O versus O/H relationship. The behavior of N/O with
increasing metallicity offers clues about the history of chemical evolution
of galaxies and the stellar populations responsible
for producing oxygen and nitrogen. 
%The location of the three 
%lensed $z \sim 2$ galaxies (S2, A2, and Cl2244-gA) are shown without 
%extinction correction (empty squares) and assuming the reddening 
%quoted for each galaxy in Table~\ref{tab:param} (filled squares). 
%
%For comparison, we show the location of 
%nearby star-forming galaxies: starburst nucleus galaxies (SBNGs) and 
%\hii\ galaxies (see Contini et al. 2002 for references). 
% 
The sample of high-redshift ($z \geq 2$) galaxies with measured N/O
abundances is still very small: the three galaxies from this paper, and MS
1512-cB58 at $z \sim 2.7$ (Teplitz et al. 2000). Surprisingly, these
four galaxies have very different locations in the N/O vs. O/H
diagram. S2 is a low-metallicity object (Z $\sim 0.04$ \zsun) 
with a low N/O ratio, similar to those
derived in the most metal-poor nearby \hii\ galaxies. In contrast, A2
is a metal-rich galaxy (Z $\sim 1.9$ \zsun) with a
high N/O abundance ratio, similar to those derived
in the most metal-rich massive SBNGs. The position of MS 1512-cB58 and 
CL2244-gA is intermediate between these two extremes showing abundance ratios
typical of low-mass SBNGs and intermediate-redshift galaxies (see also 
Lemoine-Busserolle et al. 2003, 2004).
A natural explanation for the variation of N/O at constant metallicity
is the time delay between the release of oxygen and that of nitrogen
into the ISM (e.g. Contini et al. 2002, and references therein), while
maintaining a universal IMF and standard stellar nucleosynthesis. The
``delayed-release'' model assumes that star formation is an
intermittent process in galaxies and predicts that the dispersion 
in N/O is due to the delayed release of nitrogen produced in low-mass
longer-lived stars, compared to oxygen produced in massive,
short-lived stars.
Following this hypothesis and new chemical evolution models (Mouhcine \& 
Contini 2002), we might interpret the location of S2 and A2 in the N/O 
versus O/H diagram in terms of star formation history and evolutionary stage 
of these galaxies. The low O/H and N/O abundance ratios found in S2 
might indicate a relatively young age for this object which experienced 
two or three bursts of star formation at most in the recent past. A2 seems 
on the contrary much more evolved. The location of this galaxy in the N/O 
versus O/H diagram indicates a relatively long star formation history with 
numerous powerful and extended starbursts.  

%\begin{figure}[!t]
%  \includegraphics[angle=-90,width=\columnwidth]{MS3063f5c.eps}
%  \caption{Lens-corrected velocity profiles (in \kms) of AC114$-$S2 versus 
%angular distance (in kpc), as measured on the 2D spectra from the 
%central velocities of \halpha\ (open dots) and \oiiib\ (full dots) emission 
%lines.}
%  \label{fig:rotcurve}
%\end{figure}

\begin{figure}[!t]
  \includegraphics[width=\columnwidth]{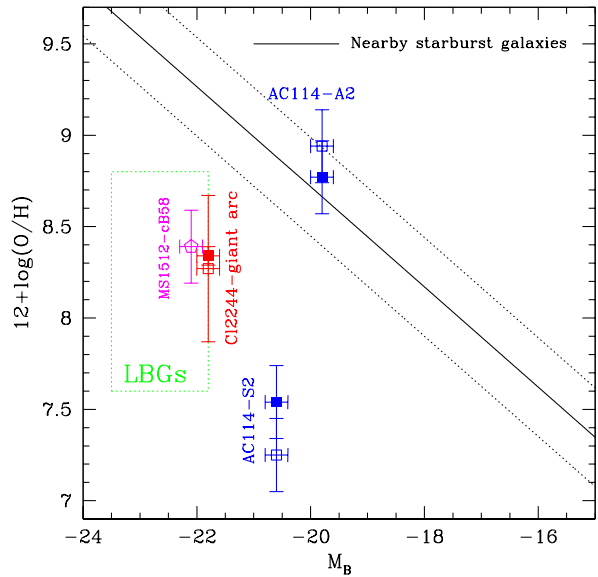}
  \caption{Oxygen abundance as a function of $B$-band absolute magnitude, 
showing the location of the three $z \sim 2$ galaxies: S2, A2 in AC 114 and the 
giant blue arc in Cl 2244, with (filled squares) and without extinction correction 
(empty squares). The metallicity--luminosity relation derived for $\sim$ 6400 
nearby ($z < 0.15$) starburst galaxies from the 2dFGRS (Lamareille et al. 2004) 
is shown as a solid line (the dashed lines indicate the typical dispersion). 
The location of high-redshift ($z \sim 3$) Lyman break galaxies (LBGs) is shown as 
a green box encompassing the range of O/H and \mabs\ derived for these objects by 
Pettini et al. (2001).}
  \label{fig:ohvsmb}
\end{figure}

We can study the evolution with redshift of the fundamental scaling relation 
between galaxy luminosity and metallicity. 
For nearby galaxies, this relation extends over $\sim 10$ magnitudes in 
luminosity and $\sim 2$ dex in metallicity (Lamareille et al. 2004 
and references therein).
%
%The high-redshift sample is best compared 
%to nearby galaxies where metallicities are derived using the same empirical strong 
%line method. Figure~\ref{fig:ohvsmb} shows the relation (solid line) derived 
%for $\sim 6400$ nearby ($z < 0.15$) star-forming galaxies from the 2dFGRS 
%(Lamareille et al. 2004), as well as high-redshift galaxies. 
%
From Figure~\ref{fig:ohvsmb}, it is immediately obvious that high-redshift 
galaxies do not conform
to today's luminosity--metallicity relation. Even allowing for the 
uncertainties in the determination of O/H, high-redshift galaxies 
have much lower oxygen
abundances than one would expect from their luminosities. This result,
already revealed by previous studies (e.g. Pettini
et al. 2001; Contini et al. 2002), is secured with the addition of the
low-luminosity and low-metallicity galaxy S2.
One interpretation of this result is that high-redshift galaxies
are undergoing strong bursts of star formation which raise their
luminosities above those of nearby galaxies with similar chemical
composition.  Another possibility is that the whole
metallicity--luminosity relation is displaced to lower abundances at
high redshifts, when the Universe was younger and the time available
for the accumulation of the products of stellar nucleosynthesis was
shorter. It should be possible to quantify this effect by measuring
metallicities in samples of galaxies at different redshifts. 
Unfortunately, to date there is no 
sample of galaxies with known chemical abundances to fill the 
gap between $z\sim 0.5$ and $z\sim 2$.
The location of A2 is more surprising. It does not follow the trend of
high-redshift objects. Instead, it lies on the luminosity--metallicity
relation of nearby objects and has the highest metallicity of any
high-redshift ($z \geq 2$) galaxy.

\begin{figure}[!t]
  \includegraphics[width=\columnwidth]{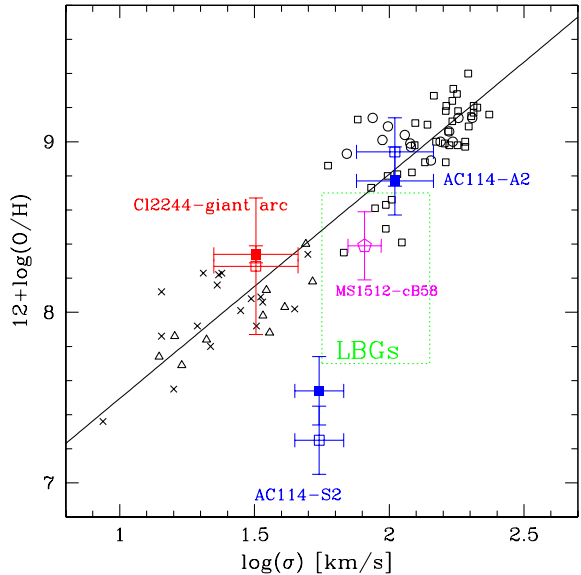}
  \caption{Oxygen abundance as a function of the
line-of-sight velocity dispersion ($\sigma$), showing the location of the three 
$z \sim 2$ galaxies: S2, A2 in AC 114 and the giant blue arc in Cl 2244, with 
(filled squares) and without extinction correction (empty squares). We show for 
comparison samples of nearby galaxies together with their linear fit 
(see Lemoine-Busserolle et al. 2003 for references). 
The location of high-redshift ($z \sim 3$) 
Lyman break galaxies (LBGs) is shown as a blue box encompassing the range of O/H and 
$\sigma$ derived for these objects by Pettini et al. (2001).}
  \label{fig:ohvssig}
\end{figure}

In Figure~\ref{fig:ohvssig} we plot the oxygen abundance against
the line-of-sight velocity dispersion ($\sigma$) for high-redshift galaxies 
and other relevant samples of nearby galaxies. This diagram
corresponds to a crude mass--metallicity sequence.  The location of
AC114-A2 is fully compatible with that of low-redshift galaxies.
However, the oxygen abundance of AC114-S2 is much smaller than the
corresponding value for low-$z$ galaxies of similar velocity
dispersions. The situation is even more dramatic if we 
consider the velocity gradient derived under the disk-like 
rotation curve hypothesis (see previous section).
Taken at face value, it would appear that the slope of
the mass--metallicity relation for galaxies with $z \geq 2$ is
different from that of low-$z$ galaxies. The behavior of Cl2244-gA 
in this diagram is peculiar; it seems that we are observing a 
nearly face-on ``disky'' galaxy as supported by its extinction close 
to zero and the very low value of $\sigma$, which would, in this case, 
not be a good tracer of its dynamics. 

\section{Implications for EMIR/GOYA project}

We have presented in this paper the firsts results of our
optical/NIR spectroscopic survey of highly magnified high-redshift 
galaxies in the core of lensing clusters, $1-2$ magnitudes fainter 
than the LBGs in the field. The results obtained on the 
physical properties of these galaxies suggest that high-$z$ objects 
of different luminosities could have quite different star formation 
histories and hence different evolutionary status. However, the number of well
observed high-$z$ objects is currently very small and larger samples
in both redshift (1.5 $\ltapprox$ z $\ltapprox$ 6) and luminosity are
required. This will be achieved with the new generation of
multi-object NIR spectrographs for the 10m-class telescopes, such as 
EMIR on the GTC. One of the main objectives of the GOYA survey 
is indeed to statistically study the physical properties of 
$z \sim 2$ galaxies.


\begin{thebibliography}
\bibitem[Bruzual \& Charlot(2003)]{2003MNRAS.344.1000B} Bruzual, G.~\& 
Charlot, S.\ 2003, \mnras, 344, 1000 
\bibitem[Colless et al.(2001)]{2001MNRAS.328.1039C} Colless, M., et al.\ 
2001, \mnras, 328, 1039 
\bibitem{} Contini, T., Treyer, M.~A., Sullivan, M., Ellis, R.~S., 2002, MNRAS, 330, 75
\bibitem{} Ebbels, T.~M.~D., Le Borgne, J.-F., Pell\'o, R., et al., 1996, MNRAS, 281, L75 
\bibitem[Erb et al.(2003)]{2003ApJ...591..101E} Erb, D.~K., Shapley, A.~E., 
Steidel, C.~C., Pettini, M., Adelberger, K.~L., Hunt, M.~P., Moorwood, 
A.~F.~M., \& Cuby, J.\ 2003, \apj, 591, 101 
\bibitem{} Kneib, J-P., Ellis, R.S., Santos, M.R., Richard, J., 2004, ApJ in press (astro-ph/0402319)
\bibitem{}Lamareille, F., Mouhcine, M., Contini, T., Lewis, I., Maddox, S. 2004, MNRAS, in press (astro-ph/0401615)
\bibitem{}Le Borgne et al., 2004, A\&A, submitted
\bibitem[Lemoine-Busserolle et al.(2003)]{2003A&A...397..839L} 
Lemoine-Busserolle, M., Contini, T., Pell{\' o}, R., Le Borgne, J.-F., 
Kneib, J.-P., \& Lidman, C.\ 2003, \aap, 397, 839 
\bibitem{} Lemoine-Busserolle, M., Contini, T., Pell{\' o}, R., Le Borgne, J.-F., 
Kneib, J.-P., \& Richard, J. 2004, \aap, submitted 
\bibitem[Lilly et al.(1995)]{1995ApJ...455...50L} Lilly, S.~J., Le Fevre, 
O., Crampton, D., Hammer, F., \& Tresse, L.\ 1995, \apj, 455, 50 
\bibitem{} Mehlert, D., Seitz, S., Saglia, R.P., et al., 2001, A\&A, 379, 96 
\bibitem[Mouhcine \& Contini(2002)]{2002A&A...389..106M} Mouhcine, M.~\& 
Contini, T.\ 2002, \aap, 389, 106 
%\bibitem{} Oey, M.~S., Kennicutt, R.~C., 1993, ApJ, 411, 137 
\bibitem{} Pell{\' o}, R., Kneib, J.-P., Le Borgne J.-F., et al., 1999, A\&A, 346, 359 
\bibitem{} Pell{\' o}ls, R., Contini, T., Lemoine-Busserolle, M., Richard, J., et al., 2003, 
in ``Gravitational Lensing: a unique tool for cosmology'' (astro-ph/0305229)
\bibitem[Pell{\' o} et al.(2004)]{2004A&A...416L..35P} Pell{\' o}, R., 
Schaerer, D., Richard, J., Le Borgne, J.-F., \& Kneib, J.-P.\ 2004, \aap, 
416, L35 
\bibitem{} Pettini, M., Kellogg, M., Steidel, C.~C., Dickinson, M.,
Adelberger, K.~L., Giavalisco, M., 1998, ApJ, 508, 539
\bibitem{} Pettini, M., Steidel, C.~C., Adelberger, K.~L., Dickinson, M., 
Giavalisco, M., 2000, ApJ, 528, 96 
\bibitem{} Pettini, M., Shapley, A.~E., Steidel, C.~C., et al., 2001, ApJ, 554, 981
%\bibitem{} Richer, M.~G., McCall, M.~L., 1995, ApJ, 445, 642
\bibitem[Schneider et al.(2003)]{2003AJ....126.2579S} Schneider, D.~P., et 
al.\ 2003, \aj, 126, 2579 
\bibitem{} Steidel, C.~C., Giavalisco, M., Dickinson, M., Adelberger, K.~L., 
1996, AJ, 112, 352 
\bibitem{} Steidel, C.~C., Adelberger, K.~L., Giavalisco, M., Dickinson, M., 
Pettini, M., 1999, ApJ, 519, 1 
\bibitem[Steidel et al.(2004)]{2004ApJ...604..534S} Steidel, C.~C., 
Shapley, A.~E., Pettini, M., Adelberger, K.~L., Erb, D.~K., Reddy, N.~A., 
\& Hunt, M.~P.\ 2004, \apj, 604, 534 
%\bibitem{} Telles, E., Terlevich, R., 1997, MNRAS, 286, 183
\bibitem{} Teplitz, H.~I., Mc Lean, I.~S., Becklin, E.~E., et al., 2000, ApJ, 533, L65
%\bibitem{} Vila Costas, M.~B., Edmunds, M.~G., 1993, MNRAS, 265, 199
%\bibitem{} Zaritsky, D., Kennicutt, R.~C., Huchra, J.~P., 1994, ApJ, 420, 87



%\bibitem{}Abbott, D. C., Bieging, J. H., Churchwell, E., \& Cassinelli, J. P.
%  1980, ApJ, 238, 196

%\bibitem{}Altenhoff, W. J., Downes, D., Pauls, T., \& Schraml, J. 1979, A\&AS, 35, 23

%\bibitem{} Arendt, R. G. 1989, ApJS, 70, 181

%\bibitem{} Assousa, G. E., Herbst, W., \& Turner, K. C. 1977, ApJ, 218, L13

%\bibitem{} Becker, R. H., White, R. L., Helfand, D. J., \& Zoonematkermani, S. 1994,
%ApJS, 91, 347

%\bibitem{} Blitz, L., Fich, M, \& Stark, A. A. 1982, ApJS, 49, 183

%\bibitem{}Bronfman, L., Nyman, L.-\AA., \& May, J. 1996, A\&AS, 115, 81

%\bibitem{} Carral, P., Kurtz, S., Rodr\'{\i}guez, L. F., Mart\'{\i}, J., Lizano, S.,
%\& Osorio, M. 1999, RevMexAA, 35, 97

%\bibitem{} Caswell, J. L., \& Lerche, I. 1979, MNRAS, 187, 201

%\bibitem{} Condon, J. J., Cotton, W. D., Greisen, E. W., Yin, Q. F.,
%Perley, R. A., Taylor, G. B., \& Broderick, J. J. 1998, AJ, 115, 1693

%\bibitem{} Finley, J. P. \& \"Ogelman, H. 1994, ApJ, 434, L25


%\adjustfinalcols
\end{thebibliography}
\end{document}